\documentclass[preprint,authoryear,12pt]{elsarticle}

\usepackage{graphics}

\usepackage{amssymb}
\usepackage{amsmath}
\journal{Advances in Space Research}

\begin{document}

\begin{frontmatter}



\title{MHD discontinuities in solar flares: continuous transitions and
       plasma heating
\tnoteref{footnote1}}


\author{Ledentsov L.S.\corref{cor}\fnref{footnote2}}
\address{Universitetsky prospekt, 13,
Moscow 119991,
Russia}
\fntext[footnote2]{P.K.\,Sternberg Astronomical Institute,
        M.V.\,Moscow State University}
\ead{koob@mail.ru}


\author{Somov B.V.\fnref{footnote3}}
\address{Universitetsky prospekt, 13,
Moscow 119991,
Russia}
\fntext[footnote3]{P.K.\,Sternberg Astronomical Institute,
        M.V.\,Moscow State University}
\ead{somov@sai.msu.ru}


\begin{abstract}

The boundary conditions for the ideal MHD equations on a plane discontinuity
surface are investigated.
It is shown that, for a given mass flux through a discontinuity, its type
depends only on the relation between inclination angles of a magnetic
field.
Moreover,
the conservation laws on a surface of discontinuity allow changing a
discontinuity type with gradual (continuous) changes in the conditions of
plasma flow.
Then there are the so-called transition solutions that satisfy
simultaneously two types of discontinuities.
We obtain all transition solutions on the basis of the complete system of
boundary conditions for the MHD equations.
We also found the expression describing a jump of internal energy of the
plasma flowing through the discontinuity.
Firstly, this allows constructing a generalized scheme of possible
continuous transitions between MHD discontinuities.
Secondly, it enables the examination of the dependence of plasma heating
by plasma density and configuration of the magnetic field near the
discontinuity surface, i.e., by the type of the MHD discontinuity.
It is shown that the best conditions for heating are carried out in the
vicinity of a reconnecting current layer near the areas of reverse
currents.
The result can be helpful in explaining the temperature distributions inside
the active regions in the solar corona during flares observed by modern
space observatories in soft and hard X-rays.

\end{abstract}


\end{frontmatter}

\parindent=0.5 cm

\section{Introduction}
\label{sec1}

%
%
Space observations of solar flares in soft and hard X-rays with spacecraft
Yohkoh and RHESSI at first revealed the characteristic structure of active
regions in the solar atmosphere related to flares
\citep{Tsuneta92,Lin-02}.
Primary energy release in a flare occurs at the top of the magnetic field
loop structure, the base of which extends below the photosphere.
Charged particles, accelerated in a flare, fall along the magnetic field
lines to the surface of the Sun and collide with the dense chromospheric
plasma.
Deceleration of the particles is accompanied by the hard X-ray emission at
the loop foot\-points
\citep{Tsuneta92}.
Chromospheric plasma heated by the collision rises along the magnetic field
lines and produces soft X-ray emission of the loop
\citep{Tsuneta96}.
Hard X-ray sources at (or above) the top of the loop are associated with the
thermal plasma heated directly in (or from) the region of primary energy
release
\citep{Masuda-94, Petrosian-02, Sui-03}.

The energy source of a flare is a non-potential part of magnetic field in
the solar corona.
The so-called free magnetic energy of interacting magnetic fluxes is
accumulated in the magnetic fields of coronal electric currents
\citep{Syrovatskii-62,Brushlinskii-80}.
Disruption or quick dissipation of such current structure can lead to the
energy release as kinetic energy of accelerated particles or thermal energy
of heated plasma.
The area of energy release is described by the reconnecting current layer
models, more exactly, the model of super-hot turbulent-current layers
\citep[][Chap. 8]{Somov-13b}.
The high-speed plasma flows form a system of discontinuous
flows near a current layer.
It is observed, for example, in numerical simulations of the magnetic
reconnection process
\citep{Shimizu-03,Ugai-05,Ugai-08,Zenitani-11}.
The MHD discontinuities are also able to convert the energy of the directed
motion of the plasma with frozen magnetic field
to the thermal energy, thus making a further
contribution to the heating of a super-hot plasma in a flare.

An abrupt change (jump) of physical characteristics of a plasma flow occurs
on the discontinuity surface
\citep{Syrovatskii-57,Anderson-63}.
It may be jumps of density and velocity of the plasma or jumps of
intensity and direction of the magnetic field lines.
The relations between these jumps determine the type of discontinuity.
Ordinary hydrodynamics allows only two types of discontinuous flows:
a tangential discontinuity and a shock wave.
However, there is much variety of possible discontinuity types in
magnetohydrodynamics (MHD) owing to the presence of the magnetic field.
Furthermore,
it is possible to transit from one flow regime to another with a gradual
(continuous) changes in the characteristics of the plasma
\citep{Syrovatskii-56,Polovin-60}.
Then one system of boundary conditions for the MHD equations on the
discontinuity surface would satisfy once two types of discontinuous flows
at the moment of transition.
This is so-called transition solution.

We aim to study the possibility of the transitions between different types
of MHD discontinuities and plasma heating by the discontinuous flows.
Initially, in Sect.~\ref{sec2},
we describe the standard classification of MHD discontinuities
\citep[e.g.,][]{Syrovatskii-57, Priest-82, Goedbloed-04, Somov-13a}
with a view to associate it with the amount of the mass flow in
Sect.~\ref{sec3}.
On this basis, we found all transition solutions for the full system of
boundary conditions
(Sect.~\ref{sec4})
and construct a demonstrative scheme of allowable transitions in MHD
(Sect.~\ref{sec5}).
Then we investigate the possibility of plasma heating by different types
of MHD discontinuities
(Sect.~\ref{sec6}).
Finally, we discuss our findings as possibly applied to calculations of the
analytical model of the magnetic reconnection in the context of the basic
physics of solar flares
(Sect.~\ref{sec7}).

%
%
\section{Boundary conditions}
\label{sec2}

We will seek a solution of the formulated problem for an MHD discontinuity,
i.e., a plasma region where the density, pressure, velocity, and magnetic
field strength of the medium change abruptly at a distance comparable to
the particle mean free path.
The physical processes inside such a discontinuity are determined by kinetic
phenomena in the plasma, both laminar and turbulent ones
\citep{Longmire-66,Tideman-71}.
In the approximation of dissipative MHD, the internal structure of a
discontinuous flow is defined by dissipative transport coefficients
(the viscosity and electric conductivity) and the thermal conductivity
\citep{Sirotina-60,Zeldovich-66}.
However, in the approximation of ideal MHD, the jump has zero thickness,
i.e., it occurs at some discontinuity surface.
%
%

We will consider a plane discontinuity surface, which is appropriate for
areas of a sufficiently small size compared to the radius of curvature of
the discontinuity surface. Let us introduce a Cartesian coordinate system
in which the observer moves with the discontinuity surface located in the
$ (y,z) $,
plane in the direction of the~$ x $
\begin{figure}
\begin{center}
\includegraphics*[width=6cm]{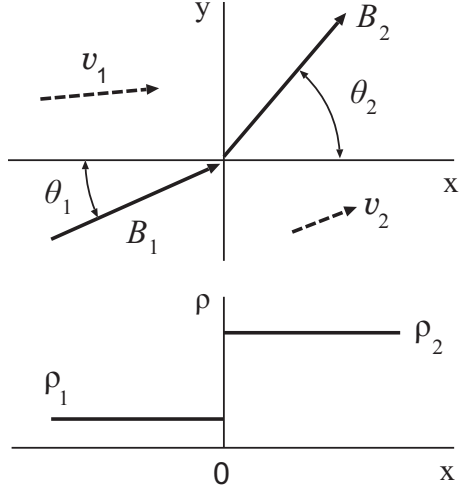}
\end{center}
\caption{Changes in the magnetic field~$ {\bf B} $, velocity
         field~${\bf v}$, and plasma density $\rho$ at the shock front
         $ x = 0 $.
         Since $ E_z \ne 0 $ in the coordinate system associated with the
         current layer, the velocity vectors are not parallel to the
         magnetic field vectors.}
\label{fig1}
\end{figure}
axis (Fig. \ref{fig1}).
In the approximation of ideal MHD, we neglect the plasma viscosity,
thermal conductivity, and electric resistivity.
The boundary conditions for the MHD equations at the discontinuity then
take the form of the following conservation laws
\citep{Syrovatskii-57}:
\begin{equation}
    \left\{ \, B_{x} \, \right\} = 0 \, ,
    \label{GRx1}
\end{equation}
\begin{equation}
    \left\{ \, \rho \, v_{x} \, \right\} = 0 \, ,
    \label{GRx2}
\end{equation}
\begin{equation}
    \left\{ \, v_{x} B_{y} - v_{y} B_{x} \, \right\} = 0 \, ,
    \label{GRx3}
\end{equation}
\begin{equation}
    \left\{ \, v_{x} B_{z} - v_{z} B_{x} \, \right\} = 0 \, ,
    \label{GRx4}
\end{equation}
\begin{equation}
    \left\{ \,
    \rho \, v_{x} v_{y} - \frac{1}{4 \pi} \, B_{x} B_{y} \,
    \right\} = 0 \, ,
    \label{GRx5}
\end{equation}
\begin{equation}
    \left\{ \,
    \rho \, v_{x} v_{z} - \frac{1}{4 \pi} \, B_{x} B_{z} \,
    \right\} = 0 \, ,
    \label{GRx6}
\end{equation}
\begin{equation}
    \left\{ \, p + \rho \, v^{\,2}_{x}
  + \frac{ B^{\, 2} }{ 8 \pi } \,
    \right\}
  = 0 \, ,
    \label{GRx7}
\end{equation}
\begin{equation}
    \left\{ \, \rho \, v_{x} \left( \frac{v^{\,2}}{2} + \epsilon
  + \frac {p}{\rho} \right)
  + \frac{1}{4 \pi} \left( \, B^{\,2} v_{x}
  - ( \, {\bf v} \cdot {\bf B} \, ) \, B_{x} \, \right) \right\}
  = 0 \, .
    \label{GRx8}
\end{equation}

\vspace{1mm}

\noindent
Here,
the curly brackets denote the difference between the values of the quantity
contained within the brackets on both sides of the discontinuity plane.
For example, Eq.~(\ref{GRx1}) implies the continuity of the normal magnetic
field component:
$$ \left\{ \, B_{x} \, \right\} =  B_{x2} - B_{x1} = 0 \, . $$
The quantities marked by the subscripts ``1'' and ``2'' refer to the side
corresponding to the plasma inflow and outflow, respectively.
%
%

In contrast to the boundary conditions in ordinary hydrodynamics, the
system of boundary conditions (\ref{GRx1})--(\ref{GRx8})
does not break up into a set of mutually exclusive groups of equations and,
hence, in principle it admits continuous transitions between different
types of discontinuous solutions as the plasma flow conditions change
continuously.
Since a smooth transition between discontinuities of various types is
possible, the local external attributes of the flow near the discontinuity
plane are taken as a basis for their classification: the presence or
absence of a mass flux and a magnetic flux through the discontinuity,
the density continuity or jump.
%
%

In the presence of transition solutions, the classification of
discontinuities in MHD can only be relative.
Indeed, a discontinuity of a given type can continuously pass into a
discontinuity of another type as the plasma inflow and magnetic field
parameters change gradually.
As it will be shown in the Sect. \ref{sec7} the type of discontinuity can
change when passing to a different point of the discontinuity surface.
In any case, since a smooth transition is possible between discontinuities
of various types, the local external signatures of the flow near the
discontinuity plane are taken as a basis for their classification:
the presence or absence of velocity, $v_x$, and the magnetic field, $B_x$,
components perpendicular to the plane (i.e., normal), the continuity or
jump in density $\rho$.
With respect to these signatures, the energy conservation law~(\ref{GRx8})
is an additional condition: at the magnetic field strength, the velocity
field, and the density jump found, Eq.~(\ref{GRx8}) defines the jump in
pressure $p$.
Thus, considering our objective of identifying the discontinuities near
the reconnecting current layer in mind, we can restrict our analysis to the
remaining seven equations: (\ref{GRx1})--(\ref{GRx7}).

It is necessary to establish conditions under which certain types of
discontinuous are formed.
Let's rotate the coordinate system about the $x$ axis in order to make
$v_{z1} = 0$.
The substitution of Eqs. (\ref{GRx1})--(\ref{GRx2}) in (\ref{GRx6}) will then
yield the equation
\begin{equation}
    { B_{x} \over {4\pi} } \, \{ B_z \} = \rho v_x \, v_{z2} \, .
    \label{GRx0}
\end{equation}

\vspace{1mm}

\noindent
Here are some simple particular solutions of this equation:
\begin{enumerate}
\item
If $B_x = 0$, $v_x = 0$, i.e., there are no magnetic flux and plasma flow
through the discontinuity, then $\{v_y\}$, $\{B_y\}$, $\{B_z\}$ and
$\{v_{z2}\}$ are arbitrary quantities as we see from Eqs. (\ref{GRx3}),
(\ref{GRx5}) and (\ref{GRx0}).
The Eq. (\ref{GRx7}) implies that the pressure and the magnetic field
strength are related by the continuity of the total pressure
\begin{displaymath}
    \left\{ \, p + \frac{ B^{\, 2} }{ 8 \pi } \,
    \right\}
  = 0 \, ,
\end{displaymath}
This solution corresponds to the classical tangential discontinuity.

\item
If $B_x = 0$, $v_x \ne 0$ (then from Eq. \ref{GRx0} $v_z = 0$), the
magnetic field is parallel to the discontinuity surface and increases with
the compression of the plasma due to a freezing condition, as we see from
Eqs.~(\ref{GRx2}), (\ref{GRx4}).
$$ \left\{ \frac{\bf B}{\rho} \right\} = 0 \, . $$
This is a perpendicular shock, that is well-known in MHD.

\item
If $B_x \ne 0$, $v_x = 0$ (then from Eq. \ref{GRx0} $\{ B_z \} = 0$),
$v_y$, $v_z$ and $B_y$ are continuous from Eqs. (\ref{GRx3}), (\ref{GRx5}).
The direction of the magnetic field does not change, however, the value of
the density may change.
This is a contact discontinuity.

\item
If $B_x \ne 0$, $v_x \ne 0$, then substitution $v_z=0$ in
Eq. (\ref{GRx0}) gives $\{ B_z \} = 0$.
As a result, Eq.~(\ref{GRx4})
is transformed into
\begin{displaymath}
    B_{z} \{ v_x \} = 0 \, .
\end{displaymath}
It allows two different solutions:

\begin{enumerate}
\item[(a)]
At the first, we consider the solution $ \{ v_x \} = 0 $.
The substitution of this condition in Eq. (\ref{GRx2}) gives Alfven wave
($\{ \rho \}=0$).
Then Eq. (\ref{GRx7}) rewriting as $\{ B_y^2 \} = 0$.

\item[(b)]
Solution $B_x = 0$ leads us to the two-dimensional picture of the
discontinuity: the velocity and the magnetic field lie in the same plane
orthogonal to the discontinuity plane.

\end{enumerate}

\end{enumerate}

Thus,
the presence or absence of the plasma overflow and penetration of the
magnetic field through the discontinuity surface allows identifying all
types of discontinuities listed above; for details of their properties see,
e.g.,
\citet{Syrovatskii-56, Somov-13a}.
The description of many discontinuities is simplified in the coordinate
system, often called the de Hoffmann-Teller system.
Such coordinate system moving along the discontinuity plane with velocity
$$ {\bf u}_{_{\, \rm HT}}
   = {\bf v} - \frac{v_x}{B_x} \, {\bf B}\, , $$
Then the vector ${\bf B}$ becomes parallel to the vector ${\bf v}$.
In particular, Alfven discontinuity describes as the rotation of the
magnetic field vector around the x axis and all oblique discontinuous
(with the magnetic field inclined to the discontinuity plane) become
two-dimensional in this case.
The de Hoffmann-Teller system is inapplicable to tangential discontinuity
and perpendicular shock wave due to the limitation $ B_x \ne 0 $, and we
will not use it.

However, taking into account that all discontinuities apart a tangential
and an Alfven discontinuity and a perpendicular shock can be reduced to
a two-dimensional form, let us begin consideration of the boundary
conditions (\ref{GRx1})--(\ref{GRx8}) with the two-dimensional case
$ 4 ({\rm b}) $.
We will study only the classification features discontinuous flows to
Sect.~\ref{sec6}.
Therefore, we will not use Eq.~(\ref{GRx8}).
Two-dimensional discontinuities will be described by the five boundary
conditions:
\begin{equation}
    \left\{ \, B_{x} \, \right\} = 0 \, ,
\quad \,\,
    \{ \rho v_x \} = 0 \, , \quad \,\,
    \left\{ \rho v_x v_y
  - {B_x B_y \over 4 \pi } \right\}
  = 0 \, ,
      \label{2d}
\end{equation}
\begin{displaymath}
    \{ v_x B_y - v_y B_x \} = 0 \, , \quad \,\,
    \left\{ \rho v_x^{\, 2} + p
  + { B_y^{\, 2} \over 8 \pi } \right\}
  = 0 \, .
\end{displaymath}

\vspace{1mm}

\noindent
In the next section we will study the change of the magnetic field on the
discontinuity surface based on the system (\ref{2d}).
%
%

%
%
\section{Inclination angles}
\label{sec3}

We will write the system of Eqs. (\ref{2d}) in the linear form with
respect to the variables~$ \{ v_x \}$, $ \{ v_y \}$, $ \{ r \}$
and~$ \{ B_y \} $:
\begin{displaymath}
    \{ v_x \} - m \, \{ r \} = 0 \, , \quad \,\,
    m \, \{ v_y \} - { B_x \over 4 \pi } \, \{ B_y \} = 0 \, ,
\end{displaymath}
\begin{equation}
    \tilde{B}_y \{ v_x \} - B_x \{ v_y \}
  + m \, {\tilde r} \,  \{ B_y\} = 0 \, ,
    \label{GRx10}
\end{equation}
\begin{displaymath}
    m \, \{ v_x \} + \{ p \}
  + {\tilde{B}_y \over 4 \pi } \, \{ B_y \} = 0 \, ,
\end{displaymath}
\noindent
where we introduced new variables $ r = 1 / \rho $ and
$ m = \rho v_x $.
The mean of the two quantities is denoted by the tilde
$ \tilde{r}~=~( r_1 + r_2 ) / 2 $.

%
%

For existing nontrivial solutions of the linear system of
Eqs.~(\ref{GRx10}), the determinant composed of its coefficients must
be equal to zero:
\vspace{1mm}
\begin{displaymath}
    \begin{vmatrix}
    \, -1 & 0      & m     & 0 \\
    0  & m      & 0     & - { B_x } / { 4 \pi } \,\, \\
    \tilde{B}_y & - B_x & 0 & m \tilde r \\
    m  & 0 & {\{ p \} } / { \{ r \}} & {\tilde{B}_y } / { 4 \pi } \,
    \end{vmatrix}
  = 0 \, .
\end{displaymath}

\vspace{1mm}

\noindent
Let us expand the determinant:
\begin{displaymath}
    {\{  p \} \over \{ r \} }
    \left( {B_x^{\, 2} \over 4 \pi } - m^{\, 2} {\tilde r}
    \right)
  + m^{\, 2}
    \left( {B_x^{\, 2} \over 4 \pi }
  + { \tilde{B}_y^{\, 2} \over 4 \pi } - m^{\, 2} {\tilde r}
    \right) =0 \, .
\end{displaymath}

\vspace{1mm}

\noindent
The latter equation imposes constraints on the admissible values of the
mass flux~$ m $:
\begin{equation}
    m^{\, 2}
  = - \, { \{ p \} \over \{ r \} } \,
    { { m^{\, 2} -{ B_x^{\, 2} } / \, { 4 \pi \, {\tilde r} } }
    \over
    { m^{\, 2}
  - \left(
    B_x^{\, 2} + \tilde{B}_y^{\, 2}
    \right)
    / \, 4 \pi \, {\tilde r} } } \, .
    \label{GRx11}
\end{equation}
%
%
We can not describe the internal structure of the discontinuity in terms
of ideal MHD.
Therefore, the discontinuity appears to the surface of zero thickness.
However, the irreversible processes in the shock waves leads to the
entropy increase.
It is essential that the amount of the entropy change does not depend on
the specific dissipation mechanism and its jump is determined by the mass,
momentum and energy conservation laws
\citep[see, e.g.,][]{Zeldovich-66}.
The entropy increase according to the Zemplen
 theorem forbids the
rarefaction shock waves.
Thus,
\begin{displaymath}
    \{ r \}
  = \frac{ 1 }{ \rho_{2} }
  - \frac{ 1 }{ \rho_{1} }
  = \frac{ \rho_{1} - \rho_{2} }{ \rho_{1} \rho_{2} }
            < 0 \, .
\end{displaymath}
\noindent
The quantity~$ m^2 $ can not be negative.
Therefore, two inequalities must hold: either
\begin{equation}
    m^{\, 2} < { B_x^{\, 2} \over 4 \pi \, {\tilde r} } \, ,
    \label{GRx13}
\end{equation}
or
\begin{equation}
    m^{\, 2} > { { B_x^{\, 2} + \tilde{B}_y^{\, 2} }
    \over { 4 \pi \, {\tilde r} } } \, .
    \label{GRx12}
\end{equation}
%
%

\vspace{1mm}

The solution of the system of Eqs. (\ref{GRx10}) is
\begin{displaymath}
    \{ v_x \}
  = C \, m
    \left( { B_x^{\, 2} \over 4 \pi } - m^{\, 2} {\tilde r}
    \right) \, ,
\end{displaymath}
\begin{equation}
    \{ v_y \}
  = C \, m { B_x \tilde{B}_y \over 4 \pi } \, ,
    \label{GRx14}
\end{equation}
\begin{displaymath}
    \{ r \}
  = C \left( { B_x^{\, 2} \over 4 \pi } - m^{\, 2} {\tilde r}
    \right) \, ,
\end{displaymath}
\begin{displaymath}
    \{ B_y \}
  = C \, m^{\, 2} \tilde{B}_y \, ,
\end{displaymath}

\vspace{1mm}

\noindent
We find the constant $ C $ by substituting the expressions
for~$ \{ v_x \} $ and~$ \{ B_y \} $
for the last equation of the system~(\ref{GRx10}):
\begin{displaymath}
   C = - \, { \{ p \} \over m^{\, 2} }
   \left( { B_x^{\, 2} + \tilde{B}_y^{\, 2} \over 4 \pi }
   -m^{\, 2} \tilde{r} \right)^{\! -1} .
\end{displaymath}
%
%

\vspace{1.5mm}

Consider the last two equations of the system~(\ref{GRx14}).
Eliminating the constant~$ C $ from them, we will obtain the
relationship between the tangential magnetic field components
\begin{displaymath}
    \{ B_y \}
  = {
    m^{\, 2} \{ r \} \over
    \left(
    { B_x^{\, 2} } / \, { 4 \pi } - m^{\, 2} {\tilde r}
    \right)
    } \,
    \tilde{B}_y \, .
\end{displaymath}

\vspace{1mm}

\noindent
Expanding the relations
$$ \{ B_y \} = B_{y2} - B_{y1} $$ and
$$ \tilde{B}_y = {1 \over 2} ( B_{y2} + B_{y1} ) \, , $$
we then have
\begin{equation}
    B_{y2}
  = {
    2 \left( { B_x^{\, 2} } / \, { 4 \pi } - m^{\, 2} {\tilde r}
      \right) + m^{\, 2} \{ r \} \over
    2 \left( { B_x^{\, 2} } / \, { 4 \pi } - m^{\, 2} {\tilde r}
      \right) - m^{\, 2} \{ r \}
    } \, B_{y1} \, .
    \label{GRx15}
\end{equation}

\vspace{1mm}

\noindent
Let us divide both parts of Eq. (\ref{GRx15}) by $ B_x $ to obtain the
relation of the angles between the magnetic field vector and the normal
to the discontinuity surface on both its sides:
\begin{displaymath}
    {\rm tan} \, \theta_2
  = {
    2 \,
    \left( \, { B_x^{\, 2} } / \, { 4 \pi } - m^{\,2} \,
      {\tilde r} \,
    \right) + m^{\,2}\, \{\, r\, \} \over
    2 \,
    \left( \, { B_x^{\, 2} } / \, { 4 \pi } - m^{\,2} \,
      {\tilde r} \,
    \right) - m^{\,2}\, \{\, r\, \}
    } \,\,
    {\rm tan} \, \theta_1 \, .
   \label{GRx9}
\end{displaymath}
Here, $ {\rm tan} \, \theta  = B_y / \,B_x $, $ r = 1  / \rho \, $,
and the tilde marks the mean values of the quantities,
$ \tilde r = (\, r_1 + r_2 \,) / \, 2 $.
Rewriting this formula by expanding the jumps $ \{ r \} $ and the means
 $ \tilde r $, we obtain
\begin{displaymath}
    {\rm tan} \, \theta_2
  = {m^{\,2} \cdot 4 \pi r_1 / B_x^{\, 2} - 1
    \over
    m^{\,2} \cdot 4 \pi r_2 / B_x^{\, 2} - 1}
    \,\, {\rm tan} \, \theta_1 \, .
   \label{GRx9.1}
\end{displaymath}
Denote
$ m_{\rm off}^{\,2} = B_x^{\, 2} / 4 \pi r_1 $ and
$ m_{\rm on}^{\,2}  = B_x^{\, 2} / 4 \pi r_{2 \,} $;
as will be shown below, $ m_{\rm off}$ and $m_{\rm on} $ correspond to
the mass flux through the switch-off and switch-on shocks.
Note that
$ m_{\rm off}~\leq~m_{\rm on} $, because according to Zemplen's theorem,
$ r_2 \leq r_1 $.
The formula for the field inclination angles takes form
\begin{equation}
   {\rm tan} \, \theta_2
 = \frac{ m^2 /m_{\rm off}^2 - 1 }{ m^2 / m_{\rm on}^2 - 1} \,
   {\rm tan} \, \theta_1 \, .
   \label{GRx9.2}
\end{equation}
%
%

Let us denote
$ m_{_{\rm A}}^{\,2} = B_x^{\, 2} / 4 \pi {\tilde r} $ and
$ m^{\,2}_\bot = \tilde{B}_y^{\, 2} / \, 4 \pi {\tilde r} $.
Since
$ r_2 \leq \tilde r \leq r_1 $, òî
$ m_{\rm off} \leq~m_{_{\rm A}} \leq m_{\rm on} $.
Then we will write conditions (\ref{GRx13}) and (\ref{GRx12}) as:
\begin{equation}
    m^{\,2} < m_{_{\rm A}}^{\, 2} \, ,
    \label{GRx10.1}
\end{equation}
\begin{equation}
    m^{\,2} > m_{_{\rm A}}^{\,2} + m_{\bot}^{\,2} \, .
    \label{GRx11.1}
\end{equation}
%
%
Note that
$ m_{\rm off} $, $ m_{\rm A} $ and $ m_{\rm on} $ are not independent but
are related by the equation
\begin{equation}
    m_{_{\rm A}}^{\, 2}
  = \frac{ 2 \,\, m_{\rm on}^{\, 2} \, m_{\rm off}^{\, 2} }
         { m_{\rm on}^{\, 2} + m_{\rm off}^{\, 2} } \, .
    \label{ma}
\end{equation}
This can be easily verified by expanding the mean~$ \tilde r $ in the
definition of $ m_{\rm A}^{\, 2} $.
%
%

Based on these results we will consider the properties of discontinuous
flows; more specifically, we will establish the possible transitions
between them.
%
%

%
%
\section{Transition solutions}
\label{sec4}

We begin to seek transition solutions with a search for the conditions of
possible transitions between various types of two-dimensional MHD flows
($ v_z = 0 $, $ B_z = 0 $), i.e., flows for which the velocity field and
the magnetic field lie in the $ (x, y) $ plane.
We call such discontinuous flows plane or two-dimensional ones.
Then, we will find the transition solutions that correspond to them and
establish the form of the solutions that are the transition ones to
three-dimensional discontinuous flows.
%
%

\begin{figure}
\begin{center}
\includegraphics*[width=11cm]{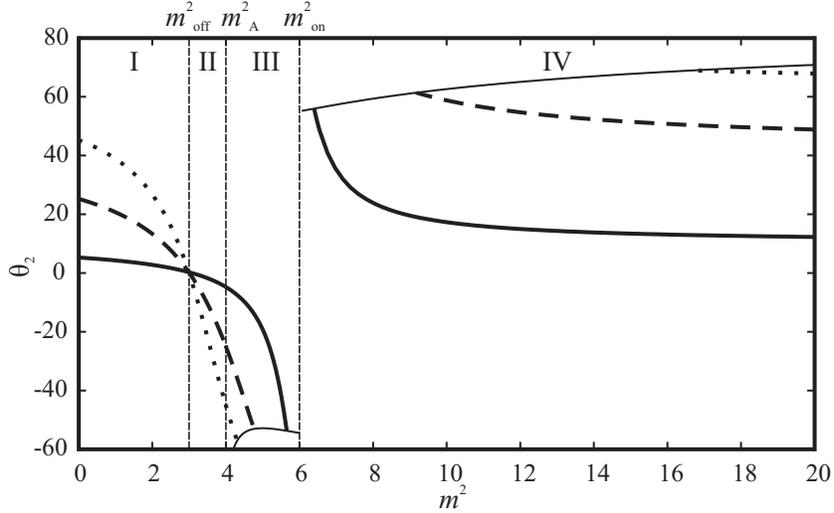}
\end{center}
\caption{Inclination angle of the magnetic field $\theta_2$ behind the
         discontinuity plane versus the square of the mass flux $m^2$ at
         various angles $\theta_1$ .
         The cases of
         $\theta_1 = 1^\circ$, $\theta_1 = 25^\circ$,
         $\theta_1 = 45^\circ$ are designated by the solid, dashed, and
         dotted lines, respectively.}
\label{fig2}
\end{figure}

Equation (\ref{GRx9.2}) along with conditions
(\ref{GRx10.1})--(\ref{GRx11.1})
describes the dependence of the magnetic field inclination angles on the
mass flux through the discontinuity.
This dependence can be specified either by the two parameters
$ m_{\rm off} $ and $ m_{\rm on} $,
or, for example, by the quantities $ \rho_1 $ and $ \{\rho\} $.
Since we are interested in the classification attributes of
discontinuities (i.e., the qualitative changes of the relation between
the angles $ \theta_1 $ and $ \theta_2 $ when varying $ m^2 $),
for the time being, we will consider Eq. (\ref{GRx9.2}) without any specific
application to certain physical conditions in the plasma.
We will choose the parameters
from clarity considerations.
Let
$ m_{\rm off}^{\,2} $, $ m_{_{\rm A}}^{\,2} $ and
$ m_{\rm on}^{\,2} $ be
related as $ 3 : 4 : 6 $.
We will measure the square of the mass flux in units of
$ m_{_{\rm A}}^{\,2} / 4 $;
then $ m_{\rm off}^{\,2} = 3 $
and
$ m_{\rm on}^{\,2} = 6 $.
%
%
%
The dependence $ \theta_2 ( m^2 ) $ is shown in Fig. \ref{fig2}
%
or three values of the angle $ \theta_1 $.
The corresponding curves behave identically.
Firstly, they intersect at one point at
$ m^2 = m_{\rm off}^{\,2} $, $ \theta_2 = 0 $ here.
Secondly,
$ \theta_2 \rightarrow - \, \theta_1 $ when
$ m^2 \rightarrow m_{_{\rm A}}^{\,2} = 4 $.
for each curve.
Thirdly, they all have a region that does not satisfy conditions
(\ref{GRx10.1}) and (\ref{GRx11.1}), located near
$ m^2 = m_{\rm on}^{\,2} $.
%
%

Let us separate out the regions in Fig.~\ref{fig2}
%
%
each of which is characterized by its own behavior of the dependence of
$ \theta_2 $ on $ m^{2} $.
In region $ {\rm I} $ ($ 0 < m^2 < m_{\rm off}^{\,2} $), the tangential
component $ B_{y2} $ of the magnetic field vector $ {\bf B}_{2} $ behind
the discontinuity surface decreases with increasing $ m^{2} $.
In this case, $ 0 < \theta_2 < \theta_1 $, i.e., when passing through the
discontinuity surface, the tangential field component weakens but remains
positive.
At $ m^{2} = m_{\rm off}^{\,2} $ when crossing the discontinuity plane,
$ B_{y2} $  becomes zero. In region~$ {\rm II} $
($ m_{\rm off}^{\,2} < m^2 < m_{_{\rm A}}^{\,2} $)
 $ B_{y2} $
is negative and increases in magnitude, but
$ - \theta_1 < \theta_2 < 0 $.
In region~$ { \rm III } $
($ m_{_{\rm A}}^{\,2} < m^2 < m_{\rm on}^{\,2} $), just as in
region~$ {\rm II} $, $ B_{y} $  changes its sign when crossing the
discontinuity plane.
Now, however, $ B_{y} $ increases in magnitude
($ \theta_2 < - \theta_1 $), remaining negative.
Finally, in region~$ {\rm IV} $ ($ m^2 > m_{\rm on}^{\,2} $)
the magnetic field is amplified ($ \theta_2 > \theta_1 $) with its
tangential component retaining the positive sign.
%
%

Thus, we have shown precisely how the behavior of the relation between the
magnetic field inclination angles and, consequently, the type of MHD
discontinuity changes with the increasing mass flux.
The regions I and II correspond to the slow shocks that, respectively, do
not reverse~($ S_{-}^{\,\uparrow} $) and
reverse~($ {S_{-}^{\,\downarrow}} $) the tangential field component.
The regions III and IV correspond to the trans-Alfven~($ {Tr} $) and
fast~($ {S_+} $) shocks, respectively.
Shock waves corresponding to the regions II and III is often called
intermediate shocks \citep{Streck-98}.
The plasma velocity is super-Alfven upstream, and sub-Alfven downstream
at the intermediate shocks while the slow shocks is everywhere sub-Alfven
and fast shocks is everywhere super-Alfven.
We use the mass flux as a classifier rather than the plasma velocity.
This classifier divides the intermediate shocks into two types which
have the different properties.
Therefore, we will use different names for the discontinuities in the
regions II and III.
The transition solutions for the discontinuities corresponding to the
adjacent regions are realized at the mass flux demarcating these regions.
%
%

Consider the behavior of the function $ \theta_2 ( m^2, \theta_1 ) $
near the boundary of regions  $ {\rm II} $ and $ {\rm III} $, where
$ m^{\,2} = m_{_{\rm A}}^{\,2} $.
The domain of definition of the function $ \theta_2 ( m^2, \theta_1 ) $
to the left and the right of $ m_{_{\rm A}}^{\,2} $ is specified by
conditions (\ref{GRx10.1}), and (\ref{GRx11.1}), respectively.
In view of Eqs. (\ref{GRx9.2}) and (\ref{ma}),
$ {\rm tan} \, \theta_2 \to - {\rm tan} \, \theta_1 $ when
$ m^{\,2} \to m_{_{\rm A}}^{\,2} $,
i.e., $ \tilde{B}_y \to 0 $.
Inequality (\ref{GRx11.1}) transforms to
$ m^{\,2} > m_{_{\rm A}}^{\,2} $.
in this case.
Therefore, the function $ \theta_2 (m^2, \theta_1) $ in region $ {\rm II} $
and region $ {\rm III} $
is defined near $ m_{_{\rm A}}^{\,2} $.
However, the right part of the condition $ m^2 $ also increases
with (\ref{GRx11.1}).
The equality
$ m^2 = m_{_{\rm A}}^{\,2} + m_{\bot}^{\,2} $ is established at some value
of $ m^2 $ in region $ {\rm III} $ and the strongest trans-Alfven shock
(increasing the magnetic energy to the greatest extent) takes place.
As the mass flux increases further, $ m^2 $ can
not satisfy conditions (\ref{GRx10.1}) and (\ref{GRx11.1})
until $ m^2 $ again
becomes equal to
$ m_{_{\rm A}}^{\,2} + m_{\bot}^{\,2} $.
This occurs in region $ {\rm IV} $, where the strongest fast shock is
observed.
%
%

Let us derive the equation of the curve bounding the function
$ \theta_2 ( m^2, \theta_1 ) $, and, hence, the strongest (for given
plasma parameters), fast and trans-Alfven shocks. Setting $ m^{\,2} $
equal to the right part of condition (\ref{GRx12}), we find
\begin{displaymath}
    B_{y1}
  = \pm \, 2 \,
    \sqrt{4 \pi \tilde r \, m^2 - B_x^2} \, - B_{y2} \, ,
    \label{max2}
\end{displaymath}
where the plus and minus correspond to regions IV and III, respectively.
Dividing the derived equation by $ B_x $, we have
\begin{equation}
   {\rm tan} \, \theta_1
 = \pm \, 2 \,
   \sqrt{ m^2 / \,m_{_{\rm A}}^{\, 2} - 1} \,
 - {\rm tan} \, \theta_2 \, .
   \label{max3}
\end{equation}
Substituting Eq. (\ref{max3}) in (\ref{GRx9.2}), we obtain the equation
of the sought-for curve
\begin{equation}
   {\rm tan} \, \theta_2 =
   \pm \,
   \frac{ m^{\,2} / \, m_{\rm off}^{\,2} - 1}{
   \sqrt{ m^{\,2} / \,m_{\rm A}^{\,2} - 1}} \, .
   \label{max}
\end{equation}
The corresponding curves are represented in Fig.~\ref{fig2}
%
%
by the thin lines.
%
%

Now, we will seek transition solutions in order of the increasing mass
flux $ m $, starting from $ m = 0 $.
Let's consider the transition between the contact discontinuity ($ {C} $)
at $ m^2 = 0 $
and the slow shock in region I.
The boundary conditions for two-dimensional discontinuities follow from
Eqs. (\ref{GRx1})--(\ref{GRx7})
when $ v_z = 0 $ and $ B_z = 0 $ are substituted:
\begin{equation}
    \left\{ \, B_{x} \, \right\} = 0 \, ,
    \quad
    \left\{ \, \rho \, v_{x} \, \right\} = 0 \, ,
    \quad
    \left\{ \,
    p + \rho \, v^{\,2}_{x} + \frac{ B_{y}^{\,2} }{ 8 \pi } \,
    \right\}
  = 0 \, ,
    \label{GRx16}
\end{equation}
\begin{displaymath}
    \left\{ \, v_{x} B_{y} - v_{y} B_{x} \, \right\} = 0 \, ,
    \quad
    \left\{ \,
    \rho \, v_{x} v_{y} - \frac{ 1 }{ 4 \pi } \, B_{x} B_{y} \,
    \right\} = 0 \, .
\end{displaymath}

\vspace{0.3mm}

\noindent
The solution of these equations in region I presented in Fig.~\ref{fig2}
%
%
corresponds to the slow shock
($ S_{-}^{\,\uparrow} $)
that does not change the sign of the tangential magnetic field component.
This can be easily verified at small $ m^2 $.
It follows from Eq.~(\ref{GRx9.2}) which, of course, remains applicable
under conditions of the two-dimensional discontinuous flows that
\begin{displaymath}
   {\rm tan} \, \theta_2 \approx
   \left( 1 -
   \frac{ m^2 }{ m_{\rm off}^2 }
 + \frac{ m^2 }{ m_{\rm on}^2  } \right) \,
   {\rm tan} \, \theta_1 \, ,
\end{displaymath}
i.e., $ 0 < \theta_2 < \theta_1 $, as it must be in the slow shock.
Moreover, when $ m^2 \rightarrow 0 $, i.e.,
$ v_x \rightarrow 0 $, it follows from the system (\ref{GRx16}) that
$ \left\{ \, v_{y} \, \right\} \rightarrow 0 $ and
$ \left\{ \, B_{y} \, \right\} \rightarrow 0 $ which is the only limiting
case for the slow shock.
%
%

It remains to show that when $ v_x \rightarrow 0 $, conditions~(\ref{GRx16})
transform to the boundary conditions at the contact discontinuity.
Indeed, the substitution of $ v_x = 0 $ in
(\ref{GRx16}) gives
\begin{equation}
    \left\{ \, B_{x} \, \right\} =
    \left\{ \, v_{y} \, \right\} =
    \left\{ \, B_{y} \, \right\} =
    \left\{ \, p \, \right\} = 0 \, ,
        \label{GRx17}
\end{equation}
At the contact discontinuity, the jump in density $ \{ \rho \} $
is
nonzero.
Otherwise, all quantities remain continuous.
Thus, solution (\ref{GRx17}) simultaneously describes the slow shock in
the limit $ v_x \rightarrow 0 $,and the contact discontinuity, i.e.,
it is the corresponding transition solution.
%
%

When crossing the boundary of regions I and II, the tangential magnetic
field component changes its sign.
The slow shock ($ S_{-}^{\,\uparrow} $) that does not reverse the
tangential field component turns into the reversing slow shock
($ {S_{-}^{\,\downarrow}} $).
The transition solution is realized on the boundary of the regions, when
$ \theta_{2} = 0 $.
The substitution of $ B_{y2} = 0 $ in (\ref{GRx16}) gives the
corresponding transition solution:
\begin{equation}
    \left\{ \, B_{x} \, \right\} = 0 \, ,
    \quad
    \left\{ \, \rho \, v_{x} \, \right\} = 0 \, ,
    \quad
    \left\{ \, p + \rho \, v^{\,2}_{x}\,
    \right\}
  = \frac{ B_{y1}^{\,2} }{ 8 \pi } \, ,
    \label{GRx22}
\end{equation}
\begin{displaymath}
    B_{x} \left\{ \, v_{y} \, \right\} = - v_{x1} B_{y1} \, ,
    \quad
    \rho \, v_{x}\left\{ \, v_{y} \,
    \right\} = - \frac{1}{4 \pi} \, B_{x} B_{y1} \, .
    \label{GRx24}
\end{displaymath}

\vspace{0.3mm}

\noindent
Eliminating $ \{ v_y \} $ from the last two equations, we find
\begin{equation}
   m^2 = {\rho_1 B_x^{\, 2} } / \, { 4 \pi } =
   m_{\rm off}^{\,2} \, ,
\end{equation}
which was to be proved.
This mass flux corresponds to the switch-off shock~($ {S_{\, \rm off}} $):
the tangential field component disappears behind the discontinuity plane.
This occurs irrespective of the angle~$ \theta_1 $ and corresponds to
the intersection of the curves at $ m^2 = m_{\rm off}^{\,2} $
in Fig.~\ref{fig2}.
%
%
%
%

The reversal of the tangential field component at the boundary of
regions II and III can be a special case of the three-dimensional
Alfven discontinuity~($ {A} $).
Since there is no density jump at the Alfven discontinuity, let us
substitute $ \{ \rho \} = 0 $ in (\ref{GRx16}):
\begin{equation}
    \left\{ \, \rho \, \right\} =
    \left\{ \, B_{x}\, \right\} =
    \left\{ \, v_{x}\, \right\} = 0 \, ,
    \quad
    \left\{ \, p + \frac{ B_{y}^{\,2} }{ 8 \pi } \,
    \right\}
  = 0 \, ,
    \label{GRx27}
\end{equation}
\begin{displaymath}
    B_{x} \left\{ \, v_{y} \, \right\}
  = v_{x} \left\{ \, B_{y} \, \right\} \, ,
    \quad
    \rho \, v_{x} \left\{ \,
    v_{y} \,
    \right\}
  = \frac{1}{ 4 \pi } \, B_{x} \left\{\, B_{y} \, \right\} \, .
    \label{GRx29}
\end{displaymath}

\vspace{0.3mm}

\noindent
If $ \{ B_y \} = 0 $, then all quantities are continuous and there is no
discontinuity. Let $ \{ B_y \} \ne 0 $.
Eliminating the ratio $ \{ B_y \} / \{ v_y \} $ from the last two equations,
we obtain
\begin{equation}
   m^{\,2} = {\rho B_x^{\, 2} } / \, { 4 \pi }
   = m_{_{\rm A}}^{\,2} \, .
   \label{mAlv}
\end{equation}
When substituting $ \{ \rho \} = 0 $ in
(\ref{GRx1})--(\ref{GRx7}), we find the
boundary conditions at the Alfven discontinuity:
\begin{equation}
    \left\{ \, \rho \, \right\} =
    \left\{ \, B_{x} \, \right\} =
    \left\{ \, v_{x} \, \right\} = 0 \, ,
    \quad
    \left\{ \, p +
    \frac{ B_{y}^{\,2 } + B_{z}^{\,2} }{ 8 \, \pi} \,
    \right\}  = 0 \, ,
    \label{GRx34}
\end{equation}
\begin{displaymath}
    B_{x} \left\{ \, v_{y} \, \right\}
  = v_{x} \left\{ \, B_{y} \, \right\} \, ,
    \quad
    \rho \, v_{x} \, \left\{ \, v_{y} \, \right\}
  = \frac{1}{4 \pi} \, B_{x} \left\{ \, B_{y} \, \right\}\, ,
\end{displaymath}
\begin{displaymath}
    B_{x} \left\{ \, v_{z} \, \right\}
  = v_{x} \left\{ \, B_{z} \, \right\} \, ,
    \quad
    \rho \, v_{x} \, \left\{ \, v_{z} \, \right\}
  = \frac{1}{4 \pi} \, B_{x} \left\{ \, B_{z} \, \right\} \, .
    \label{GRx36}
\end{displaymath}

\vspace{0.3mm}

\noindent
Comparison of the systems (\ref{GRx27}) and (\ref{GRx34}) shows that the boundary
conditions (\ref{GRx27}) describe the transition discontinuity between
the slow shock in the limit
$ \{ \rho \} \rightarrow 0 $ and the Alfven flow at
$ v_z = 0 $ and
$ B_z= 0 $.
The discontinuity is a special case of the Alfven discontinuity that
reverses the tangential magnetic field component.
Trans-Alfven discontinuities reverse and enhance the tangential field
component.
They occupy region III and are adjacent to the Alfven mass flux (\ref{mAlv})
on the right side.
The conditions for the transition to the Alfven discontinuity are identical
to (\ref{GRx27}).
%
%

\begin{figure}
\begin{center}
\includegraphics*[width=10cm]{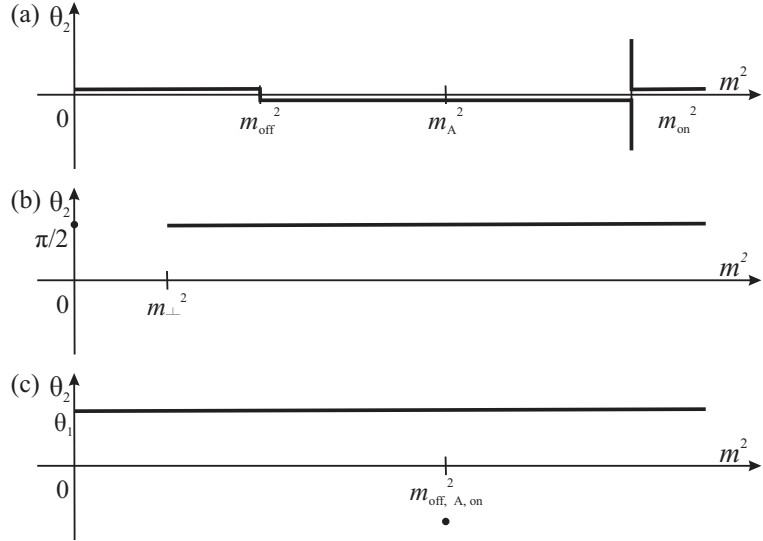}
\end{center}
\caption{Schematic behavior of the dependence $\theta_2(m^2)$
         at $\theta_1 = 0$ (a), $B_x = 0$ (b), and $ \{\rho \} = 0 $ (c).}
\label{fig3}
\end{figure}

There can be no flow near the boundary of regions~III and~IV in some range
of mass  fluxes.
For this reason, the transition between the trans-Alfven and fast
shock is forbidden.
The range narrows as the initial
inclination angle of the magnetic field decreases to
$ \theta_1 = 0 $
(Fig. \ref{fig3}).
%
The strongest fast shock takes place at the
minimum possible mass flux that is admissible by condition (\ref{GRx11.1}).
As the mass flux increases, the tangent of the field
inclination angle behind the discontinuity plane
decreases, asymptotically approaching
$ {\rm tan} \, \theta_2 =( {\rho_2 / \rho_1} ) \,
  {\rm tan} \, \theta_1 $.
%
%

%
%
\section{The scheme of transitions}
\label{sec5}

Varying $ \rho_1 $, $ \{ \rho \} $ and $ B_x $ leads to contraction or
extension of the curves presented in Fig.~\ref{fig2}
%
%
along the coordinate axes without any change of their overall structure.
For zero $ \theta_1 $, $ B_x $ and $ \{ \rho \} $ the behavior of
dependence $ \theta_2 ( m^{2} ) $ is shown in Fig.~\ref{fig3}.
%
%
In view of Eq. (\ref{GRx9.2}) when $ \theta_1 \rightarrow 0 $, the
angle $ \theta_2 $ also approaches zero at almost all values of $ m^2 $
(the case of $ m^2 = m_{\rm on}^{\,2} $
will be considered separately).
If $ B_{y1} = 0 $, then $ B_{y2} = 0 $
(Fig.~\ref{fig3}{\it a}).
%
%
In this case, the boundary conditions for the two-dimensional discontinuities
(\ref{2d}) take the form
\begin{equation}
    \left\{ \, \rho \, v_{x} \, \right\} = 0 \, ,
    \quad
    \left\{ \, p + \rho \, v^{2}_{x} \,
    \right\}
  = 0 \, ,
    \label{GRx39}
\end{equation}
\begin{displaymath}
    \left\{ \, B_{x} \, \right\} = 0 \, ,
    \quad
    \left\{ \, v_{y} \, \right\} = 0 \, ,
    \label{GRx41}
\end{displaymath}

\vspace{0.3mm}

\noindent
which corresponds to an ordinary hydrodynamic shock that propagates
according to the conditions $ B_{y1} = 0 $ and $ B_{y2} = 0 $
along the magnetic field.
System (\ref{GRx39}) is the transition solution between the oblique shocks
in the limit $ \theta_1 \rightarrow 0 $ and the parallel
shock~($ {S_\parallel} $).
%
%

As the angle $ \theta_1 $ decreases, the discontinuity between the
admissible mass fluxes for the fast and the trans-Alfven shocks will also
decrease.
Conditions (\ref{2d}) in this case give
\begin{equation}
    \left\{ \, B_{x} \, \right\} = 0 \, ,
    \quad
    \left\{ \, \rho \, v_{x} \, \right\} = 0 \, ,
    \quad
    \left\{ \, p + \rho \, v^{\,2}_{x} \,
    \right\}
  = - \frac{ B_{y2}^{\,2} }{ 8 \pi } \, ,
    \label{GRx43}
\end{equation}
\begin{displaymath}
    B_{x} \left\{ \, v_{y} \, \right\} = v_{x2} B_{y2} \, ,
    \quad
    \rho \, v_{x} \left\{ \, v_{y} \,
    \right\}
    = \frac{1}{ 4 \pi } \, B_{x} B_{y2} \, .
    \label{GRx45}
\end{displaymath}

\vspace{0.3mm}

\noindent
From the simultaneous solution of the last two equations, we have
\begin{equation}
    m^2
  = { \rho_2 B_x^{\, 2} / \, 4 \pi } =
    m_{\rm on}^{\,2} .
\end{equation}
For this mass flux, Eq.  (\ref{GRx9.2})
has no unique solution. A
nonzero $ \theta_1 $ can correspond to zero $ \theta_2 $.
The tangential
magnetic field component appears behind the shock
front, corresponding to the switch-on shock~($ { S_{\, \rm{on}} } $).
It is indicated in Fig. \ref{fig3}{\it a}
%
%
by the vertical segment at
$ m^2 = m^2_{\, \rm{on}} $.
The switch-on shock can act as the transition one
for the trans-Alfven and fast shocks in the limit
$ \theta_1 \rightarrow 0 $, but this, of course, requires that
$ m^2 \rightarrow m^2_{\, \rm{on}} $.
Otherwise, there will be the transition to the parallel shock according
to (\ref{GRx39}).
%
%

To establish the form of the transition solution between the parallel shock
and the contact discontinuity, we will set $ v_x = 0 $ in (\ref{GRx39}).
After that we will obtain
\begin{equation}
    \left\{ \, B_{x} \, \right\} =
    \left\{ \, v_{y} \, \right\} =
    \left\{ \, p \, \right\} = 0 \, ,
    \label{GRx49}
\end{equation}

\vspace{0.3mm}

\noindent
This system of equations corresponds to the contact discontinuity
(\ref{GRx17}), orthogonal to the magnetic field lines.
It describes the transition discontinuity between the parallel shock in
the limit $ v_x \rightarrow 0 $ and the contact discontinuity.
Such a transition takes place at $ m^2 = 0 $ in Fig.~\ref{fig3}{\it a}.
%
%
%
%

When $ B_x = 0 $ (Fig.~\ref{fig3}{\it b}),
%
%
$ m^2_{\, \rm{on}} $ becomes zero and all nonzero mass fluxes locate in
region IV in Fig.~\ref{fig2}.
%
%
To find the boundary conditions corresponding to them, let us substitute
$ B_x = 0 $ in (\ref{2d}).
We obtain
\begin{equation}
    \left\{ \, \rho \, v_{x} \, \right\} = 0 \, , \quad
    \left\{ \,
    p + \rho \, v^{\,2}_{x} + \frac{ B_{y}^{\,2} }{ 8 \pi }
    \right\}
  = 0 \, ,
    \label{GRx51}
\end{equation}
\begin{displaymath}
    \left\{ \, v_{x} B_{y} \, \right\} = 0 \, , \quad
    \left\{ \, v_{y} \, \right\} = 0 \, .
    \label{GRx53}
\end{displaymath}

\vspace{0.3mm}

\noindent
These conditions characterize a compression shock propagating
perpendicularly to the magnetic field. In the general case of a
perpendicular shock~($ {S_\bot} $) we will find the boundary conditions
by substituting $ B_x = 0 $ in (\ref{GRx1})--(\ref{GRx7}):
\begin{displaymath}
    \left\{ \, \rho \, v_{x} \, \right\} = 0 \, ,
    \quad
    \left\{ \,
    p + \rho \, v^{\,2}_{x}
      + \frac{B_{y}^{\,2} + B_{z}^{\,2}}{8 \pi} \,
    \right\}
  = 0 \, ,
    \label{GRx59}
\end{displaymath}
\begin{equation}
    \left\{ \, v_{x} B_{y} \, \right\} = 0 \, ,
    \quad
    \left\{ \, v_{y} \, \right\} = 0 \, ,
    \label{GRx58}
\end{equation}
\begin{displaymath}
    \left\{ \, v_{x} B_{z} \, \right\} = 0 \, ,
    \quad
    \left\{ \, v_{z} \, \right\} = 0 \, .
\end{displaymath}

\vspace{0.3mm}

\noindent
Equations (\ref{GRx51}) are then the boundary conditions for the transition
discontinuity between the fast shock in the limit $ B_x \rightarrow 0 $ and
the perpendicular shock with the magnetic field directed along the
$ y $-axis.
This transition can take place only at mass fluxes that satisfy
inequality (\ref{GRx11.1}) which takes the form $ m^{\,2} > m^{\,2}_\bot $ at
$ B_x = 0 $
(Fig.~\ref{fig3}{\it b}).
%
%
%
%

To determine the boundary conditions for the discontinuity at
$ m^2 = 0 $ (Fig.~\ref{fig3}{\it c}),
%
%
let us substitute $ B_x = 0 $ and
$ v_x = 0 $ in (\ref{GRx1})--(\ref{GRx7}).
In this case, the magnetic field and the velocity field are parallel to
the discontinuity surface and can undergo arbitrary jumps in magnitude and
direction, while the jump in pressure is related to the jump in the
magnetic field by the condition
\begin{equation}
    \left\{ \,
    p + \frac{ B_{y}^{\,2} + B_{z}^{\,2} }{ 8 \pi} \,
    \right\}
  = 0 \, .
    \label{GRx63}
\end{equation}

\vspace{0.3mm}

\noindent
This corresponds to the tangential discontinuity~($ {T} $).
The contact discontinuity, the slow shock, and the Alfven discontinuity
can pass to it in the limit $ B_x \rightarrow 0 $ under certain conditions.
Let us find the corresponding transition solutions.
Firstly, let us substitute $ B_x = 0 $ in the boundary conditions for the
contact discontinuity (\ref{GRx17}).
We will obtain the transition solution
\begin{equation}
    \left\{ \, B_{y} \, \right\} =
    \left\{ \, v_{y} \, \right\} =
    \left\{ \, p \, \right\} = 0 \, .
    \label{GRx62}
\end{equation}

\vspace{0.3mm}

\noindent
It describes the tangential discontinuity (\ref{GRx63}) for the zero field
component $ B_z $ in the absence of jumps
$ \{ v_y \} $ and $ \{ B_y \} $.
%
%
%
Secondly, the conditions for the oblique shocks
(\ref{2d}) at $ B_x = 0 $ and $ v_x = 0 $ are the transition solution
\begin{equation}
    \left\{ \, p + \frac{ B_{y}^{\,2} }{ 8 \pi } \,
    \right\}
  = 0 \, ,
    \label{GRx61}
\end{equation}

\vspace{0.3mm}

\noindent
that corresponds to the plane tangential discontinuity (\ref{GRx63}) at
$ B_z = 0 $.
Finally, the boundary condition for the Alfven discontinuity (\ref{GRx34})
after the substitution of $ B_x = 0 $ gives the transition solution
\begin{equation}
    \left\{ \, \rho \, \right\} = 0 \, , \quad
    \left\{ \, p + \frac{ B_{y}^{\,2} + B_{z}^{\,2} }{ 8 \pi } \,
    \right\}
  = 0 \, ,
    \label{GRx60}
\end{equation}

\vspace{0.3mm}

\noindent
that describes the tangential discontinuity (\ref{GRx63}), with
out any jump in density $ \rho $.
%
%

The angle $ \theta_2 $ of the strongest trans-Alfven shock is
defined by Eq. (\ref{max3}).
In view of (\ref{max3}), $ \theta_2 \rightarrow - \theta_1 $ when
$ B_x \rightarrow 0 $.
Thus, the trans-Alfven shocks degenerate into a special case of the
Alfven discontinuity as the magnetic flux decreases.
Of course, the density jump can also be set equal to zero for any type of
flow.
In this case, all parameters
$ m^2_{\, \rm{off}} $, $ m_{_{ \rm{A} }}^{\, 2} $ and
$ m^2_{\, \rm{on}} $ will be equal to the flux (\ref{mAlv}), at which the
Alfven discontinuity (\ref{GRx27}) takes place.
In Fig. \ref{fig3}{\it c},
%
%
it is denoted by $ m^2_{\, \rm{off,A,on}} $.
At other mass fluxes, the differences in plasma characteristics on
different sides of the discontinuity will disappear; the discontinuity will
be absent as such.
%
%

\begin{figure}
\begin{center}
\includegraphics*[width=10cm]{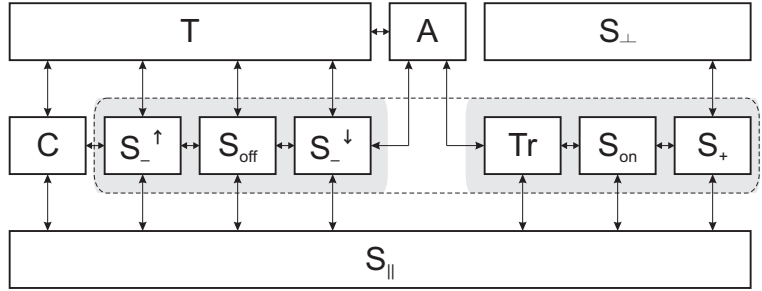}
\end{center}
\caption{Scheme of continuous transitions between MHD discontinuities.
         The dashed line encircles the set of discontinuities corresponding
         to the block of oblique shocks in the scheme by
         \citet{Syrovatskii-56}.
         The shading inside the contour highlights the "slow" (left) and
         "fast" (right) components of the scheme proposed by
         \citet{Somov-94}.
}
\label{fig4}
\end{figure}

Let us combine the properties of discontinuous solutions systematized above
into the scheme of permitted transitions shown in Fig.~\ref{fig4}.
%
%
Here, the two-dimensional discontinuities are located in the middle row in
order of increasing the mass flux and the three-dimensional discontinuities
are located in the upper row.
The one-dimensional parallel shock~($ S_{\, \parallel} $)
occupies the lower row.

\begin{figure}
\begin{center}
\includegraphics*[width=4cm]{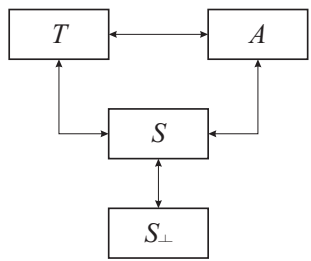}
\end{center}
\caption{Scheme of continuous transitions between
         MHD discontinuities \citep{Syrovatskii-56}}
\label{fig5}
\end{figure}

The first description of transition solutions
\citep{Syrovatskii-56}
contained only four types of discontinuous flows: a tangential
discontinuity~($ T $) and Alfven~($ A $), oblique~($ S $) and
perpendicular~($ S_{\bot} $) shocks (Fig. \ref{fig5}).
The corresponding scheme of continuous transitions between discontinuous
solutions of the equations of ideal MHD showed such transitions to be
possible in principle, but it was definitely incomplete.
Firstly, it did not have some of the discontinuous solutions, in
particular, the parallel shock~($ S_{\, \|} $) and the contact
discontinuity~($ C $).
Secondly, the block of oblique shocks ($ S $) combined several different
discontinuities at once: fast~($ S_+ $) and slow~($ S_- $) shocks,
switch-on~($ S_{\, \rm on} $)
and switch-off ~($ S_{\, \rm off} $), shocks, and trans-Alfven ($ Tr $),
shocks, the possibility of transitions between which requires a separate
consideration.
Subsequently, this picture of transitions was supplemented based on the
correspondence between the shocks and small amplitude waves
\citep{Somov-94}.
\begin{figure}
\begin{center}
\includegraphics*[width=7cm]{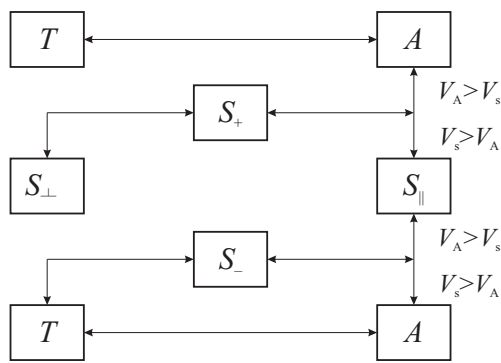}
\end{center}
\caption{Scheme of continuous transitions between
         MHD discontinuities \citep{Somov-94}}
\label{fig6}
\end{figure}
Although this approach allows the possible transitions (Fig. \ref{fig6})
and even their conditions to be correctly specified, it provides no
description of the specific form of transition solutions between the
discontinuities under consideration.
%
%

We group individual elements for the convenience of comparing the
generalized scheme of transitions with those proposed previously.
Syrovatskii's scheme
\citep{Syrovatskii-56}
is consistent with Fig. \ref{fig4}
%
%
if we combine the elements~$ {S_{-}^{\,\downarrow}} $,
$ {S_{\, \rm{off}}} $, $ {S_{-}^{\,\uparrow}} $, $ {Tr} $,
$ {S_{\, \rm{on}}} $ and
$ {S_{+}} $ into one block of ``oblique shocks''~($ {S} $), while omitting
the question of whether any transitions inside the block are possible and
disregard the contact discontinuity~($ {C} $) and the parallel
shock~($ {S_{\, \parallel}} $).
The scheme proposed in
\citep{Somov-94},
includes the parallel shock~($ {S_{\, \parallel}} $) and the separation of
oblique shocks into the slow one ($ {S_{-}} $), corresponding to
condition (\ref{GRx10.1}), and the fast one~($ {S_+} $)
corresponding to condition (\ref{GRx11.1}).
It is quite obvious that the scheme of transitions we propose is a proper
and natural generalization of the two previous schemes.
Our scheme contains not only evolutionary types of discontinuities but
also non-evolutionary ones: the switch-on, Alfven, and trans-Alfven shocks.
%
%

%
%
\section{Jump in internal energy}
\label{sec6}

To determine the plasma heating efficiency, let us turn to the boundary
condition (\ref{GRx8}), which is the energy conservation law
\citep{Ledentsov-13}.
Using Eq. (\ref{GRx2}), we will find the jump in internal energy
from (\ref{GRx8})
\begin{equation}
      \left\{ \, \epsilon \, \right\}
  = - \left\{ \frac{ v^{\,2} }{2} \right\}
    - \frac{1}{m} \left\{ \, v_{x} p \, \right\}
    - \frac{1}{ 4 \pi m }
      \left\{ \, B^{\,2} v_{x}
    - ( \, {\bf v} \cdot {\bf B} \, ) \, B_{x} \, \right\} .
      \label{GRx65}
\end{equation}
Using the mean velocities $ \tilde v_{x} $,
$ \tilde v_{y} $ and $ \tilde v_{z} $, we will write the first term as
\begin{displaymath}
    - \left\{ \frac{v^{\,2}}{2} \right\}
  = - \tilde v_{x} \left\{ v_{x} \right\}
    - \tilde v_{y} \left\{ v_{y} \right\}
    - \tilde v_{z} \left\{ v_{z} \right\}.
      \label{GRx66}
\end{displaymath}
We will express the jumps in tangential velocity components in terms of
the jumps in tangential magnetic field components using
Eqs.~(\ref{GRx1}), (\ref{GRx2}),
(\ref{GRx5}) and (\ref{GRx6}) as
\begin{displaymath}
    \left\{ v_{y} \right\}
  = \frac{ B_x }{ 4 \pi m } \left\{ B_{y} \right\} , \quad
    \left\{ v_{z} \right\}
  = \frac{ B_x }{ 4 \pi m } \left\{ B_{z} \right\} .
    \label{GRx67}
\end{displaymath}
Now, the first term on the right side of Eq. (\ref{GRx65})
appears as
\begin{equation}
    - \left\{ \frac{v^{\,2}}{2} \right\}
  = - \tilde v_{x} \left\{ v_{x} \right\}
    - \frac{ \tilde v_{y} B_{x} }{ 4 \pi m} \, \{ B_{y} \}
    - \frac{ \tilde v_{z} B_{x} }{ 4 \pi m} \, \{ B_{z} \} .
    \label{GRx68}
\end{equation}
%
%

Let us substitute in the second term written as
 \begin{displaymath}
     - \frac{1}{m} \left\{ v_{x} p \right\}
   = - \frac{ \tilde p }{ m } \{ v_{x} \}
     - \frac{ \tilde v_{x} }{ m } \{ p \} ,
     \label{GRx69}
 \end{displaymath}
 %
the pressure jump from Eq. (\ref{GRx7}), namely
 \begin{displaymath}
     \left\{  p \right\}
   = - m \{ v_{x} \}
     - \frac{ \tilde B_{y} }{ 4 \pi } \, \{ B_{y} \}
     - \frac{ \tilde B_{z} }{ 4 \pi } \, \{ B_{z} \} .
     \label{GRx70}
 \end{displaymath}
 %
 Here, as in Eq. (\ref{GRx65}), we made use of condition
 (\ref{GRx2}).
 As the result, the second term in Eq. (\ref{GRx65}) takes on the form
 \begin{equation}
     - \frac{ 1 }{ m } \left\{ v_{x} p \right\}
   = - \frac{ \tilde p }{ m } \, \{ v_{x} \}
     + \tilde v_{x} \{ v_{x} \}
     + \frac{ \tilde v_{x} \tilde B_{y} }{ 4 \pi m } \, \{ B_{y} \}
     + \frac{ \tilde v_{x} \tilde B_{z} }{ 4 \pi m } \, \{ B_{z} \} .
     \label{GRx71}
 \end{equation}

 Let us now open the scalar product
 $ ( \, {\bf v} \cdot {\bf B} \, ) $ in the third term:
 \begin{displaymath}
     - \frac{ 1 }{ 4 \pi m }
       \left\{ \, B^{\,2} v_{x}
     - ( \, {\bf v} \cdot {\bf B} \, ) \, B_{x} \,
       \right\}
   =
 \end{displaymath}
 \begin{displaymath}
   - \frac{ 1 }{ 4 \pi m }
       \left\{
       ( v_x B_y - v_y B_x ) B_y
     + ( v_x B_z - v_z B_x ) B_z
       \right\} .
     \label{GRx73}
 \end{displaymath}
 %
 Then, let us apply conditions
Eq. (\ref{GRx3}) and Eq. (\ref{GRx4}) to the derived equation:
 \begin{displaymath}
     - \frac{ 1 }{ 4 \pi m }
       \left\{ \, B^{\,2} v_{x}
     - ( \, {\bf v} \cdot {\bf B} \, ) \, B_{x} \,
       \right\}
   =
 \end{displaymath}
 \begin{equation}
   - \frac{ v_x B_y - v_y B_x }{ 4 \pi m } \, \{ B_y \}
     - \frac{ v_x B_z - v_z B_x }{ 4 \pi m } \, \{ B_z \} .
     \label{GRx74}
 \end{equation}
 %
Thus,
each of the three terms in the right side of Eq. (\ref{GRx65}) is expressed
via individual jumps of the normal velocity components and tangential
magnetic field components.
Let us substitute Eqs. (\ref{GRx68})--(\ref{GRx74}) in (\ref{GRx65})
\begin{displaymath}
      \left\{ \, \epsilon \, \right\}
  = - \frac{ \tilde p }{ m } \, \{ v_{x} \}
    + \frac{ \tilde v_{x} \tilde B_{y}
           - \tilde v_{y} B_{x} }{ 4 \pi m } \, \{ B_{y} \}
    - \frac{ v_x B_y - v_y B_x }{ 4 \pi m } \, \{ B_y \} \, +
\end{displaymath}
\begin{displaymath}
  + \, \frac{ \tilde v_{x} \tilde B_{z}
         - \tilde v_{z} B_{x} }{ 4 \pi m } \, \{ B_{z} \}
  - \frac{ v_x B_z - v_z B_x }{ 4 \pi m } \, \{ B_z \} .
    \label{GRx75}
\end{displaymath}
This equation can be simplified if we expand the means appearing in it:
 \begin{displaymath}
       \left\{ \, \epsilon \, \right\}
   = - \frac{ \tilde p }{ m } \, \{ v_{x} \}
     - \frac{ \{ v_x \} \{ B_y \} }{ 16 \pi m } \, \{ B_y \}
     - \frac{ \{ v_x \} \{ B_z \} }{ 16 \pi m } \, \{ B_z \} .
     \label{GRx76}
 \end{displaymath}
Factoring out $ - \{v_x\} / m = - \{r\} $, we derive the final equation,
which expresses the internal energy jump at the discontinuity through the
jumps of the inverse density and tangential magnetic field components
\begin{equation}
      \left\{ \, \epsilon \, \right\}
  = - \{ r \}
      \left( \tilde p
    + \frac{ \{ B_y \}^2 + \{ B_z \}^2 }{ 16 \pi }
      \right) .
    \label{GRx77}
\end{equation}
%
For two-dimensional discontinuities, it takes quite a simple form,
\begin{equation}
    \left\{ \, \epsilon \, \right\}
  = - \{ r \}
    \left( \tilde p + \frac{ \{ B_y \}^2 }{ 16 \pi }
    \right) .
    \label{GRx78}
\end{equation}
%
%

Equation (\ref{GRx77}) allows definitive conclusions regarding the change
in plasma internal energy when crossing the discontinuity surface to be
reached.
First, the internal energy increases, because, according to Zemplen's
theorem, $ - \{ r \} > 0 $ and $ \tilde p $ and $ \{ B_y \}^2 $ are
positive.
Second, the change in internal energy consists of two parts: the
thermodynamic and magnetic ones.
The latter depends on the magnetic field configuration and, hence, on the
type of discontinuity.
Let us express the tangential magnetic field components in
Eq.~(\ref{GRx78}) in terms of the corresponding inclination angles:
\begin{displaymath}
    \left\{ \, \epsilon \, \right\}
  = - \{ r \} \, \tilde p
    - \{ r \} \, \frac{B_x^{\, 2}}{ 16 \pi }
    \left( {\rm tan} \, \theta_2
         - {\rm tan} \, \theta_1 \right)^{\, 2} .
\end{displaymath}
Then, we will take the thermodynamic part of the heating independent of the
type of discontinuity as the zero point and will measure the jump in
internal energy itself in units of $ - \{ r \} {B_x^{\,2}}/\,{ 16 \pi } $.
For this purpose, let us make the substitution
\begin{displaymath}
    \left\{ \, \epsilon \, \right\}^{\, \prime}
  = -  \frac{ 16 \pi }{ \{ r \} B_x^{\, 2} }
    \left( \left\{ \, \epsilon \, \right\} + \{ r \} \tilde p
    \right)\, .
\end{displaymath}
We will obtain the equation
\begin{equation}
    \left\{ \, \epsilon \, \right\}^{\, \prime}
  = \left( {\rm tan} \, \theta_2 - {\rm tan} \, \theta_1
    \right)^{\, 2} \, .
    \label{GRx79}
\end{equation}

\begin{figure}
\label{fig7}
\begin{center}
\includegraphics*[width=11cm]{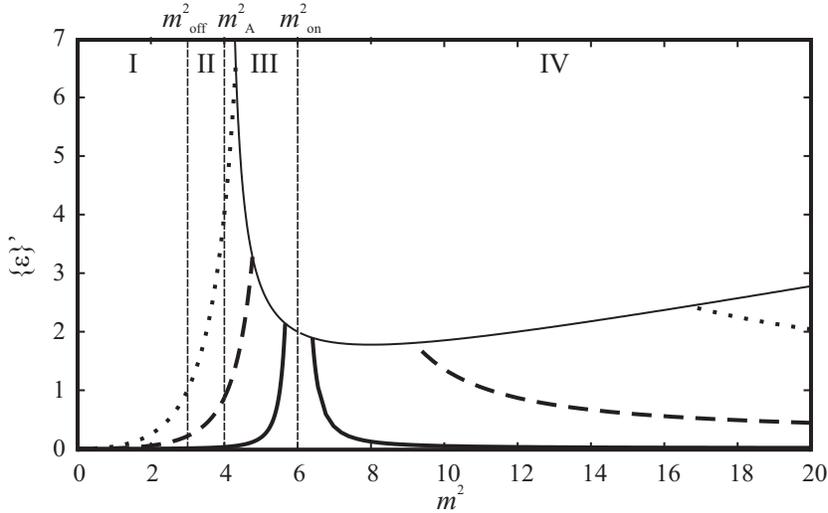}
\end{center}
\caption{Jump in internal energy $\epsilon$ versus mass flux for various
         angles $\theta_1$: $5^\circ$ (solid lines),
         $25^\circ$(dashed lines), and $45^\circ$ (dotted lines).}
\end{figure}

The magnetic part of the internal energy jump depends on the magnetic field
configuration and, as a result, on the discontinuity type.
For the calculations shown in Fig. \ref{fig2}, the dependencies of the
internal energy jump on the mass flow passing through the discontinuity
which were calculated using formula (\ref{GRx78}), are shown in
Fig. \ref{fig7}.
Here, zero stands for the thermodynamic part of the jump, which does not
depend on the mass flow.
It can be seen that, at the specified plasma parameters, the maximum jump
of the internal energy is induced by the strongest trans-Alfven shock wave;
moreover, its magnitude rapidly grows with an increase in the magnetic
field incidence angle $\theta_1$.
The correlations between the efficiency of plasma heating by other types
of discontinuities depend on specific conditions of the medium.
For example, heating by slow shock waves can be both lower than heating
with fast shock waves at smaller $\theta_1$, and higher than that at
bigger $\theta_1$.
In any case, the heating depends on the shock strength. The larger the
change in magnetic energy density, the higher the temperatures to which
the plasma will be heated.

%
The curve describing the jumps in internal energy at the strongest
trans-Alfven and fast shocks is
\begin{equation}
    \left\{ \, \epsilon \, \right\}^{\, \prime} =
    \frac{ \left(m^{\, 2} / \, m_{\rm off}^{\, 2}
    - \, m^{\, 2} / \, m_{\rm on}^{\, 2} \right)^2 }
         { m^{\,2} / \,m_{_{\rm A}}^{\,2} - 1 } \, .
\label{GRx80}
\end{equation}
It is indicated by the thin line in Fig. \ref{fig7}.
%
%

\vspace{1mm}

%
%
%
%
\section{The model of reconnection}
\label{sec7}

\begin{figure}
\begin{center}
\includegraphics*[width=5.6cm]{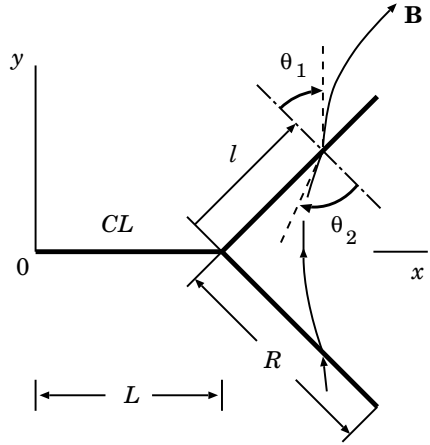}
\end{center}
\caption{The configuration of electric currents (thick straight segments)
         consists of a current layer ($ CL $) and four segments
         (discontinuity surfaces) of finite length $R$ attached to its
         endpoints; $L$ is the current layer half-width
         \citep{Bezrodnykh-07}.}
\label{fig8}
\end{figure}

\citet{Markovskii-89}
suggested a two-dimensional stationary model of magnetic reconnection,
that generalizes and combines the famous reconnection models by
\citet{Petschek-64}
and
\citet{Syrovatskii-71}.
Later on,
\citet{Bezrodnykh-07}
constructed a two-dimensional analytical model of reconnection in a plasma
with a strong magnetic field that included a thin current layer ($CL$) and
four discontinuous MHD flows of finite length $R$ attached to its endpoints.
$L$ is a halfwidth of the current layer.
Figure \ref{fig8} displays the configuration of discontinuities inside
which the electric currents flow perpendicular to the figure plane.
The normal magnetic field component vanishes on the current layer and is
equal to a given constant $\beta$ on the MHD shock waves inclined to
the $x$-axis at angle $\alpha$.
The field at infinity grows linearly with a proportionality
coefficient~$ \gamma $.
The quantities $L$, $R$, $\alpha$, $\beta$, and $\gamma$ are free
parameters of this model used below.

Magnetic field can be written as
$ B = ( \, B_x(x,y), \, B_y(x,y), \, 0 \, )$, i.e., only two of its
components, $B_x$ and $B_y$, are nonzero and depend only on the $x$ and
$y$ coordinates.
It follows from the magnetic field potentiality that the function
$ B = B_x - i B_y $ is an analytic function of variable $ z = x + i y $
in the exterior of the sections on the complex plane shown in
Fig.~\ref{fig8}.
Thus,
the model considered in this section is reduced to finding the analytic
function $B = B_x - iB_y$ from the linear combination of $B_x$ and $B_y$
specified at its boundary.
Since this problem is symmetric with respect to the $x$ and $y$ axes,
it will suffice to consider it only in the first quadrant.
Then the boundary condition for the function $B$ is formulated as follows:
$ B_y = 0 $ on the current layer and surface ($ x = 0 $);
$ B_x \, {\rm sin} \, \alpha + B_y \, {\rm cos} \, \alpha = - \beta $
on the discontinuity surface; $ B_x = 0 $ on the surface ($ y = 0, x > L $)
and $ B = i \gamma z + o(1) $ at $ z \to \infty $.
\citet{Bezrodnykh-07} constructed the sought-for analytic function in an
explicit form by the method of the conformal mapping of the domain outside
the system of sections onto the half-plane.
%
\begin{figure}
\begin{center}
\includegraphics*[width=10cm]{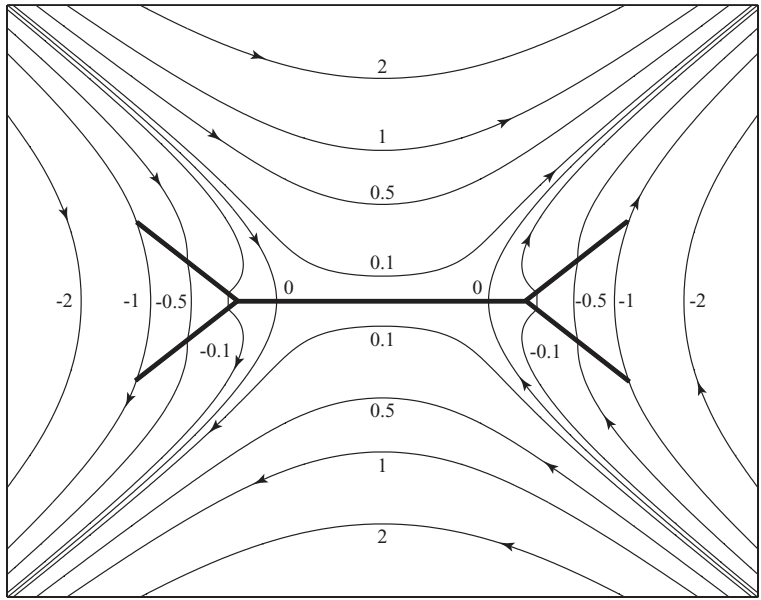}
\end{center}
\caption{Magnetic field lines (thin curves) with an indication of the
         values of the vector potential. The regions of direct and reverse
         currents inside the current layer (the thick horizontal straight
         segment) are seen.
         The magnetic field undergoes a jump on the attached discontinuity
         surfaces (inclined straight segments)
         \citep{Bezrodnykh-07}.}
\label{fig9}
\end{figure}
%
\begin{figure}
\begin{center}
\includegraphics*[width=7cm]{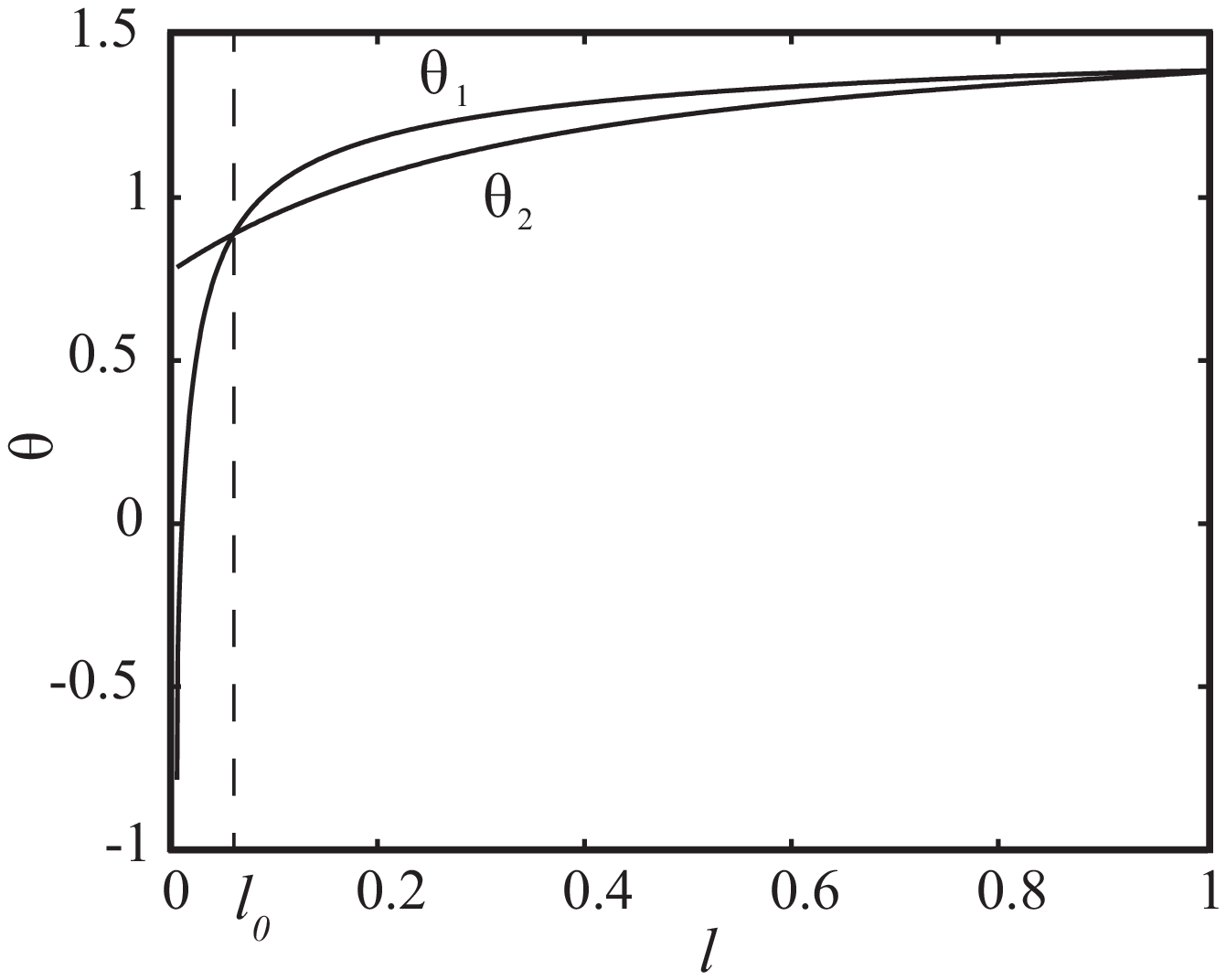}
\end{center}
\caption{Distribution of the angles $\theta_1$ and $\theta_2$ at the
         attached shock.
         The variable $l$ is the distance measured from the current layer
         endpoint along the discontinuity surface \citep{Bezrodnykh-11}.}
\label{fig10}
\end{figure}

Figure \ref{fig9}
shows magnetic field lines calculated in the special case of reconnection
($L = 1$, $R = 1$, $\alpha=45^\circ$, $\beta=1$, $\gamma=1$) that can be
taken as a basis for comparison with many other possible reconnection
regimes.
The figure demonstrates that the result of solving the problem in the main
(central) part of the reconnection region is a current layer ($CL$).
It is intersected by two symmetrically located magnetic field lines;
the points of intersection
\citep[for more details on their properties, see][]{Syrovatskii-71,
Somov-13b} separate the current layer areas with the field circulation
relative to them having opposite signs.
Thus,
the Syrovatskii-type current layer that consists of a direct current and
two attached reverse currents is actually located in the central part of
the reconnection region.

In specific astrophysical applications, in particular, in solar flares,
the model of the so-called "super-hot" $ (T_e > 10\,\, {\rm keV}) $
turbulent-current layer
\citep{Somov-13b}
should be used to determine the physical parameters of this region.
The advantage of an analytical model is the possibility to investigate
more general patterns that do not depend on the detailed assumptions of a
physical reconnection model.
Consider some of the properties of discontinuous flows in the vicinity of
a current layer predicted by the analytical model of
\citet{Bezrodnykh-07}.

A magnetic field calculation in the approximation of a strong field near
the current structure gives the angles of incidence and refraction of the
field on the discontinuity surfaces from which the types of discontinuities
are determined.
The characteristics of the plasma flowing through the discontinuity change
with the distance measured from the point of attachment to the current
layer along the discontinuity surface.
The set of these characteristics determines the type of discontinuity in
each point of the attached surface.
Moving from point to point along the surface of the discontinuity we will
see the change of the flow conditions, and hence the change of the type of
discontinuity.
The different types of discontinuous MHD solutions will correspond to the
different regions of the attached surface.
The transition solutions will realize at the boundaries between these
regions.

For example, Fig. \ref{fig10} taken from \citet{Bezrodnykh-11} shows how
the magnetic field inclination angles change with coordinate at the
discontinuity for one specific calculation.
The discontinuity surface can be arbitrarily divided into three regions
in accordance with the types of flows.
The trans-Alfven shock is immediately adjacent to the current layer.
This is the domain of negative angles $\theta_1$ in Fig. \ref{fig10}.
As one recedes from the current layer, a fast shock is observed
($\theta_2 > \theta_1$) up to the point of intersection between the plots
of $\theta_1$ and $\theta_2$ at $ l = l_0 $.
The discontinuity ends with a slow shock ($\theta_2 < \theta_1$) that
gradually passes into a continuous flow ($\theta_2 = \theta_1$) at
$ l = 1 $.
The absolute values of the angles $\theta_1$ and $\theta_2$ tend to the
same value near the current layer ($ l = 0 $ in Fig. \ref{fig10}).
Therefore, the Alfven discontinuity can be realized at this point.
We also expect further continuous transitions at $ l = 0 $ in accordance
with the scheme of transitions (Fig. \ref{fig4}) but the accuracy of the
Fig. \ref{fig10} does not allow us to say it for sure.

As the plasma flow parameters change continuously, the type of shock must
change through transition discontinuities.
The first transition occurs between the trans-Alfven and fast shocks in
the case we consider.
As has been shown above, the switch-on wave (\ref{GRx43}) must act as a
transition discontinuity.
Indeed, as can be seen from Fig. \ref{fig10}, the magnetic field at the
point of the transition between the trans-Alfven and fast shocks is normal
to the discontinuity surface ($\theta_1 = 0$) in the inflow and has a
tangential component ($\theta_2 > 0$) in the outflow.
This transition cannot be made only by a gradual change in the mass flux
through the discontinuity.
The angle of incidence of the magnetic field should be simultaneously
reduced to reduce the discontinuity between the admissible mass fluxes
for the fast and trans-Alfven shocks.
In addition, both trans-Alfven and fast shocks satisfy inequality
(\ref{GRx11.1}).
Hence, for the parameters of the medium to change smoothly, the mass flux
through the switch-on wave must also satisfy this inequality:
\begin{displaymath}
    m_{\rm off}^{\,2} > m_{\rm A}^{\,2} + m_{\bot}^{\,2} \, ,
\end{displaymath}
or
\begin{displaymath}
    { B_x^{\, 2} \over 4 \pi r_2} >
    { B_x^{\, 2} + \tilde{B}_y^{\, 2} \over 4 \pi {\tilde r} } \, .
\end{displaymath}
Simplifying this relation, we obtain a constraint on the possible
inclination of the magnetic field behind the discontinuity plane:
\begin{equation}
    {\rm tg}^{\,2} \, \theta_2 <
    \frac { 2\,\{ \, \rho \, \} }{ \rho_1 } \, .
\end{equation}

The second transition between the fast and slow shocks at $l=l_0$ cannot
be made by a continuous change in the mass flux through the discontinuity.
The intersection of the curves in Fig. \ref{fig10} suggests that there is
no jump in magnetic field strength at the point separating the fast and
slow shocks.
Only the contact discontinuity can correspond to such a field structure.
However, as has been shown above, the ratio of the tangents of the
angles $ {\rm tan} \theta_2 / {\rm tan} \theta_1 $ in the fast shock
asymptotically tends only to $\rho_2 / \rho_1$ from above and can take
on unity typical of a contact discontinuity only at $\{ \rho \} = 0$.
This, along with the boundary conditions (\ref{GRx17}), gives the absence
of a discontinuity at the point of contact between the fast and slow
shocks.
The jumps in density, velocity, magnetic field, and pressure at this point
must be zero.

Thus, the discontinuity surface turns out to be physically separated into
two regions: the inner part consists of trans-Alfven and fast shocks,
while the outer part is a slow shock.
When changing the initial model parameters, we can obtain a reconnection
regime in which the discontinuity ends with a fast shock, while the outer
part of the discontinuity is absent altogether.
This suggests that the inner part of the discontinuity is attributable to
the reconnection process itself and is closely related to the presence of
a reverse current at the endpoints of the current layer, which was clearly
shown by \citet{Bezrodnykh-07}, while the outer part depends strongly on
the factors affecting the overall topology of current layers: the presence
or absence of a ``magnetic obstacle'', non-uniformity of the plasma
distribution in the reconnection region.

We can trace some analogies of our conclusions with the results of
present-day numerical MHD simulations of fast reconnection
\citep{Shimizu-03, Ugai-05, Ugai-08, Zenitani-11}.
As a result of reconnection, the plasma being ejected from a current layer
is gathered into the so-called ``plasmoid'' separated from the surrounding
plasma by a system of slow shocks.
The structure and intensity of the latter depend on the plasmoid sizes and
density.
In addition, the characteristics of the outer part of the discontinuity
surface in \citet{Bezrodnykh-11} depend on the model geometry.
Closer to the current layer, in the region of reverse currents, the
numerical models give a complex system of shocks.
Since the derived trans-Alfven waves are non-evolutionary, the structure
of the inner part of the discontinuity described by
\citet{Bezrodnykh-07, Bezrodnykh-11} must also become more complicated
and can have similarities to the results of numerical simulations.
On the whole, the interrelationship between the process of magnetic
reconnection and the formation of a system of accompanying discontinuous
flows requires a further comprehensive study.

Equation (\ref{GRx77}) allows us to draw some definite conclusions
concerning variations in the internal plasma energy upon the passage
though the discontinuity surface.
First, the internal energy increases since, according to the Zemplen's
theorem, $-\{ r \} > 0$, and $ \tilde p $ and $ \{ B_y \} ^2 $ are
positive quantities.
Second, the internal energy change is made of two parts, namely, a
thermodynamic part, which is defined by the plasma pressure and a magnetic
one, which is related to variations in the magnetic field structure in the
vicinity of the discontinuity surface.

\begin{figure}
\begin{center}
\includegraphics*[width=8cm]{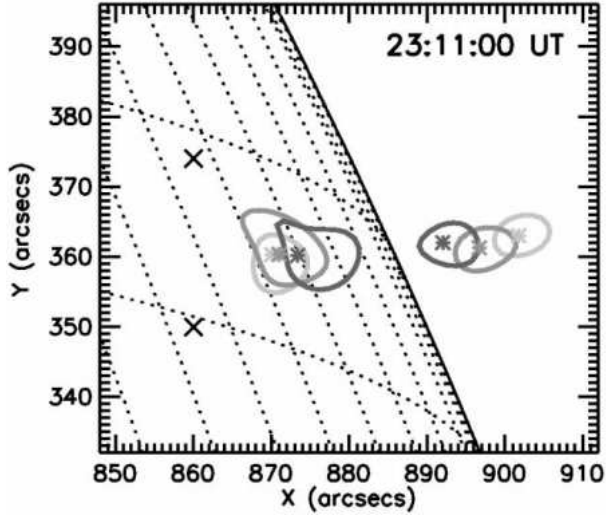}
\end{center}
\caption{RHESSI images in different energy bands at 23:11:00 UT.
         The three contours ($80\%$ of the peak flux in each image) on the
         solar disk indicate the top of loops in the energy bands
         (contour line shade from light to dark): 6--8, 10--12, and
         16--20 keV.
         The contours ($80\%$ of the peak flux of the coronal source in
         each image) above the limb are for the (from light to dark)
         10--12, 12--14, and 14--16 keV bands.
         The asterisk marks the centroid of each source.
         The crosses mark the two footpoints of the X-ray loop
         \citep{Sui-03}.}
\label{fig11}
\end{figure}

The maximum heating at discontinuities can be expected in a magnetic field
with a sharply varying geometry.
These conditions are met in the region of magnetic reconnection.
Similar discontinuity structures can be introduced as independent elements
into analytical reconnection model.
Earlier we showed that the waves that are near the face ends of the current
layer (where reverse current are generated) are trans-Alfven shock waves.
Moving away from the current layer, the jump of the magnetic field
intensity at the discontinuity decreases and the density jump vanishes.
Thus,
the best conditions for plasma heating occur near the region of reverse
currents.
However, this conclusion is purely qualitative.
Since the input parameters of the model
\citet{Bezrodnykh-07, Bezrodnykh-11}
are some abstract quantities, we can not produce any numerical estimates
about the real plasma heating.
This process requires further study.
Nonetheless, heating by shock waves can help form a super-hot plasma
observed with modern X-ray space observatories.
\citep[Fig. \ref{fig11},][]{Sui-03}.

\section{CONCLUSIONS}
\label{sec8}


The correspondence between the standard classification of discontinuous
flows in ideal MHD and the characteristic parameter of plasma flow -- the
value of the mass flow through the discontinuity -- is set up.
On this basis, all of allowed transition solutions are found.
As a result, the generalized scheme of the continuous transitions between MHD discontinuities is built.
It contains the discontinuities that were not represented in the earlier
schemes, such as the contact discontinuity, swich-on and swich-off shocks.
Some types of discontinuous flows, for example the trans-Alven shock waves,
are non-evolutionary.
They are also included in the generalized scheme of transitions.

Within a framework of the simplified analytical model of the magnetic
reconnection various areas of connected to the current layer discontinuity
surface are identified with different types of MHD shock waves.
In particular, regions of the trans-Alven shock waves are located near the
ends of the current layer (in the presence of reverse currents).
Division of the attached to the current layer discontinuity surfaces into
two areas is found.
That division is result of the different origin of discontinuities.
The quasi-stationary inner region is due to the reverse currents in the current
layer, and the outer region is caused mainly by the boundary conditions of
the magnetic reconnection process and by the speed of reconnection.

We have considered dependence of the plasma heating as the thermodynamic
parameters of plasma as the type of MHD discontinuity.
The larger the jumps in plasma density and magnetic energy density at the
discontinuity, the stronger the heating.
Such conditions occur near a region of the reverse currents in the reconnection process.
We believe that this result will be useful for quantitative explaining the
temperature distributions of the super-hot plasma in solar flares observed
by modern space X-ray observatories.

\vspace{0.5mm}

%
%
\section{ACKNOWLEDGEMENTS}
\label{sec9}

We are grateful to the referees for very helpful remarks.
This work was supported by the Russian Foundation for Basic Research
(project no. 14-02-31425-mol-a).


\end{document}